# Canonical Form of Lyapunov's Second Method in Mathematical Modelling and Control Design

Myroslav K. Sparavalo[1]

[1] NYC Transit, NY, NY, USA; National Technical University of Ukraine, Kiev, Ukraine
`mksparavalo@yahoo.com`

**Abstract.** The objective of the paper is to put canonical Lyapunov function (CLF), canonizing diffeomorphism (CD) and canonical form of dynamical systems (CFDS), which have led to the generalization of the Lyapunov's second method, in perspective of their high efficiency for Mathematical Modelling and Control Design. We show how the symbiosis of the ideas of Henri Poincaré and Nikolay Chetaev leads us to CD, CFDS and CLF. Our approach successfully translates into mathematical modelling and control design for special two-angles synchronized longitudinal maneuvering of a thrust-vectored aircraft. The essentially nonlinear five-dimensional mathematical model of the longitudinal flight dynamics of a thrust-vectored aircraft in a wing-body coordinate system with two controls, namely the angular deflections of a movable horizontal stabilizer and a turbojet engine nozzle, is investigated. The wide-sense robust and stable in the large tracking control law is designed. Its core is the hierarchical cascade of two controlling attractor-mediators and two controlling terminal attractors embedded in the extended phase space of the mathematical model of the aircraft longitudinal motion. The detailed demonstration of the elaborated technique of designing wide-sense robust tracking control for the nonlinear multidimensional mathematical model constitutes the quintessence of the paper.

## 1 Introduction

The development of a general method of construction of Lyapunov functions for specific solutions of specific systems of ordinary differential equations has been a very stubborn problem for a long time. The front attack on it has proved to be ineffective. That is why the basic idea to try to solve the problem backwards or from behind has been undertaken. What does it mean? The front attack suggests to try to construct Lyapunov function for the original system that stays unchangeable for all the course of actions. But we can do it conversely. Assume that we have already some standard or canonical Lyapunov function (CLF) common for all systems. The next logical step can be to bring each of them to its specific standard (canonical) form by some transformation so that the one and canonical Lyapunov function bound together to be subject to the Lyapunov theorem. The transformation can be obtained through the first integrals of the systems. Thus, the basic idea profits from two other ideas, which have been cultivated for many years. First, this is CLF. One should be noted that the idea to use CLF has been exploited for decades but the limited success was achieved only for some specific classes of simple nonlinear systems and never as certain general approach. See, for example, [1]-[3]. Second, the Chetaev's idea of utilizing first integrals of systems to construct Lyapunov functions. The Chetaev's idea has also received big attention for elaboration and implementation. See [4] - [7].

Technically, the solution to the problem of developing the general method of construction of Lyapunov functions has been found via introducing the concept of canonizing diffeomorphism (CD), canonical form of dynamical systems (CFDS) and canonical Lyapunov function (CLF), where the latter is the positive definite diagonal quadratic form [8]. The classic formulation of Lyapunov theorem of asymptotical stability, in which each different system requires different Lyapunov function to find, is replaced with the generalized one. It has introduced CLF that serves all the dynamical systems as some kind of one-for-all tester of stability. Fig. 1 illustrates the concept.

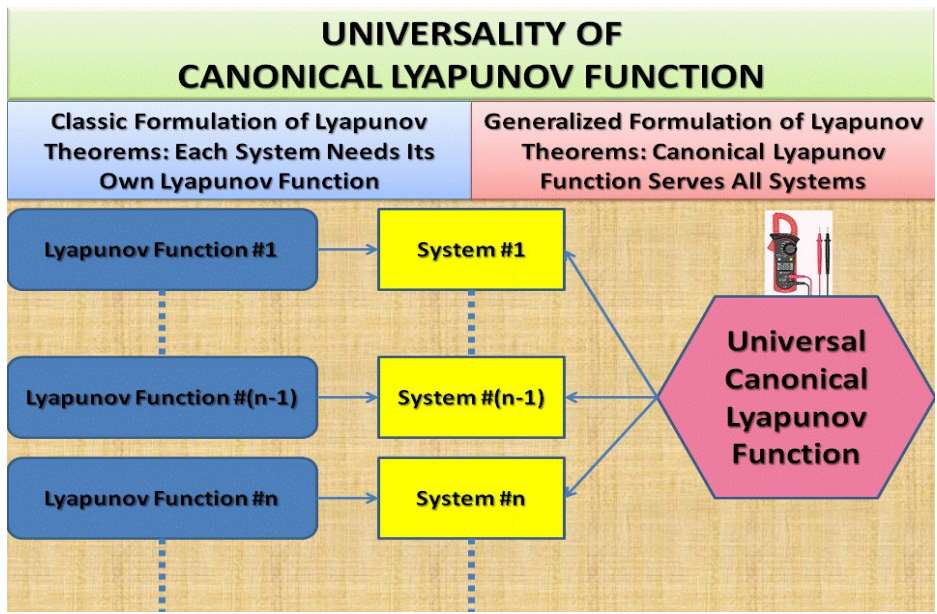

**Fig. 1.** Canonical Lyapunov function as a one-for-all tester for the stability of dynamical systems.

The mainstay of this theoretical approach is built on three backbones. First, this is the cascade of sequential flattening diffeomorphims $\theta = \theta_1 \circ ... \circ \theta_n$ resulting to CD. The last one shapes up the canonical form of the dynamical system under study, which is the second one. The canonical form of Lyapunov functions is the third backbone. This technique has allowed to develop the general procedure of the investigation of the system stability using the concept of Lyapunov functions. The idea to solve the problem of stability in the canonical form is aimed at finding the most effective way of extracting as much information of the system stability hidden in the right-hand sides of its differential equations as possible. For this purpose, the three-point Poincaré's strategy of the investigation of differential equations and manifolds has been used.

## 2   Sketching out the use of CD, CFDS and CLF

Let the asymptotical stability of an integral curve $x_t = M_{x_1} \cap M_{x_2}$ of the 2D non-autonomous system $S$ in its original form $S_x = \left\{\dfrac{dx}{dt} = f(t,x)\right\}$ be studied. Here $\{M_{x_i}\}_{i=1}^{i=2} \subset R^3_{t,x_1,x_2}$ are 2D invariant manifolds. The canonizing diffeomorphism $\bar{\varphi}(\bar{\varphi}^{-1})$ transforms $S_x$ into CFDS $S_y = \left\{\dfrac{dy}{dt} = f^2(t,y)\right\}$, $\{M_{x_i}\}_{i=1}^{i=2}$ into the invariant manifolds $\{M_{y_i} = \{y_i = 0\}\}_{i=1}^{i=2}$ and $x_t$ into the integral curve $y_t = (t, y \equiv 0)$. The information of the stability of the system $S$ stored in the right hand-sides of the differential equations of its canonical form $S_y$ can be retrieved and used with CLF $V = y_1^2 + y_2^2$. To transfer the property of stability from the integral curve $y_t$ of $S_y$ to the one $x_t$ of $S_x$, the condition of convergency must be valid [8].

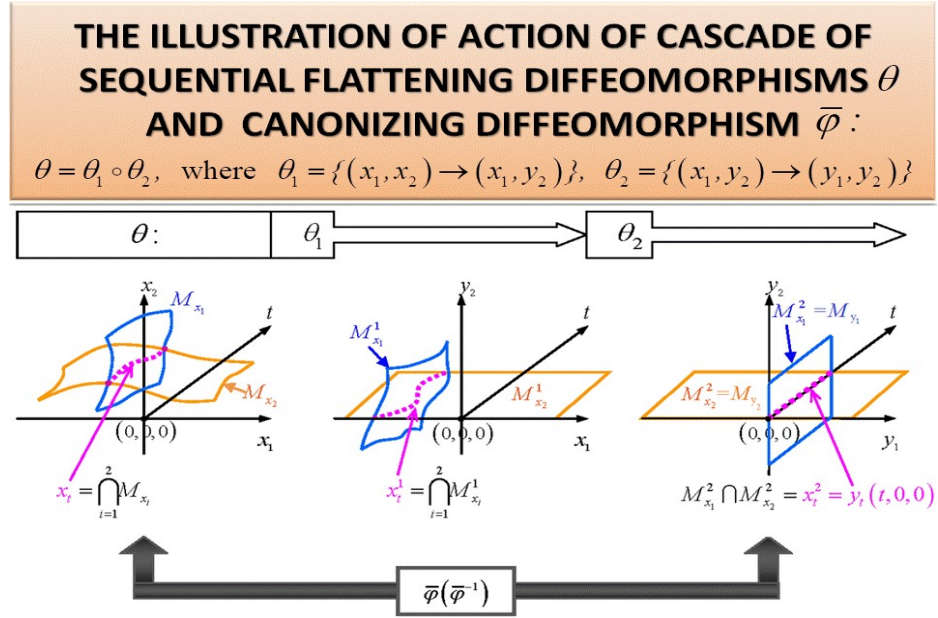

**Fig. 2.** The diffeomorphism $\bar{\varphi}(\bar{\varphi}^{-1})$ flattens invariant manifolds, rectifies integral curves and translates them in a straight line starting from the origin point $(0,0,0)$.

For dynamical systems with control the latter gives us considerable latitude to make dynamical systems to assume the canonical form and to use CLF. To illustrate how to profit from this idea we will consider an elementary example. Let us have some physical system $S$ that in the space $\mathbb{R}^3_{t,x_1,x_2}$ can be described in the following form of 2D dynamical system $S_x = \left\{\dfrac{dx_1}{dt} = f_1(x_2) + u_1, \; \dfrac{dx_2}{dt} = f_2(x_1) + u_2\right\}$. Our aim is to find the feedback control that transmutes a given curve $L_x = \{x_1 = \chi_1(t), \; x_2 = \chi_2(t)\}$ into an asymptotically stable solution of $S_x$. If we successively apply to the system $S$ in $S_x$-form two following transformations $\theta = \theta_1 \circ \theta_2$, where $\theta_1 = \{x_1 = y_1 + \chi_1(t), \; x_2 = x_2\}$, $\theta_2 = \{x_2 = y_2 + \chi_2(t), \; y_1 = y_1\}$, we will get CFDS $S_y$ of $S$ in the space $\mathbb{R}^3_{t,y_1,y_2}$

$$S_y = \left\{\begin{array}{l} \dfrac{dy_1}{dt} = f_1(y_2 + \chi_2(t)) + u_1 - \dfrac{d\chi_1(t)}{dt} = F_1(t, y_2, u_1) \\ \dfrac{dy_2}{dt} = f_2(y_1 + \chi_1(t)) + u_2 - \dfrac{d\chi_2(t)}{dt} = F_2(t, y_1, u_2) \end{array}\right\}.$$

Actually, $\bar{\varphi}(\bar{\varphi}^{-1}) = \{x_1 = y_1 + \chi_1(t), \; x_2 = y_2 + \chi_2(t)\}$ is canonizing diffeomorphism (CD) for $S_x$-form of $S$. But the controls $(u_1, u_2)$ are free variables and if we introduce two functions $g_1(y_1)$ and $g_2(y_2)$, which we will endue with special quality later, compose the equations $\{F_1(t, y_2, u_1) = g_1(y_1), \; F_2(t, y_1, u_2) = g_2(y_2)\}$ and solve them for $(u_1, u_2)$ correspondingly, we will receive

$$\left\{u_1 = g_1(y_1) - f_1(y_2 + \chi_2(t)) + \dfrac{d\chi_1(t)}{dt}, \; u_2 = g_2(y_2) - f_2(y_1 + \chi_1(t)) + \dfrac{d\chi_2(t)}{dt}\right\}.$$

The system $S$ takes on CFDS $S_y = \left\{\dfrac{dy_1}{dt} = g_1(y_1), \; \dfrac{dy_2}{dt} = g_2(y_2)\right\}$ due to the control laws constructed above that makes arbitrary curve $L_x$ integral one for $S_x$. Now let $\left\{g_1(0) = 0, \; \dfrac{dg_1(y_1)}{dy_1} < 0 \forall y \in \mathbb{R}\right\}$ and $\left\{g_2(0) = 0, \; \dfrac{dg_2(y_2)}{dy_2} < 0 \forall y \in \mathbb{R}\right\}$. The two last conditions guarantee that CLF $V_y = y_1^2 + y_2^2$ provides the integral curve $L_x$ of $S_x$-form of the system $S$ with asymptotical stability and it becomes one-dimensional attractor. In the space $\mathbb{R}^3_{t,x_1,x_2}$ for the original $S_x$-form of the system $S$ the control $(u_1, u_2)$ assumes the following shape

$$\left\{u_1 = g_1(x_1 - \chi_1(t)) + \dfrac{d\chi_1(t)}{dt} - f_1(x_2), \; u_2 = g_2(x_2 - \chi_2(t)) + \dfrac{d\chi_2(t)}{dt} - f_2(x_1)\right\}.$$

We can clearly see that both of the above-written feedback control laws contain the components, $-f_1(x_2)$ and $-f_2(x_1)$, which cancel the right-hand sides of the corresponding equations of $S_x$-form, while the rest components are responsible for shaping the asymptotical stability of the curve $L_x$. The demonstrated modus operandi is the loose interpretation of the method of control design based on the topological techniques from the standpoint of Lyapunov functions developed in the early 1990s [9].

## 3   Flight dynamics of a thrust-vectored aircraft: mathematical modelling and wide-sense robust control design by CLF

The five-dimensional mathematical model of the longitudinal flight dynamics of a thrust-vectored aircraft in a wing-body coordinate system is governed by

$$\begin{cases} \dfrac{dv}{dt} = \dfrac{P}{m}\cdot\cos(\alpha+\delta_P) - \dfrac{0.5\rho v^2 S C_X}{m} - g\sin\theta = A_1(\bullet), \\ \dfrac{d\theta}{dt} = \dfrac{P}{mv}\sin(\alpha+\delta_P) + \dfrac{0.5\rho v S C_Y}{m} - \dfrac{g}{v}\cdot\cos\theta = A_2(\bullet), \\ \dfrac{d\alpha}{dt} = q - A_2(\bullet) \Leftrightarrow \left\{\dfrac{d\vartheta}{dt} = q, \alpha = \vartheta - \theta\right\}, \\ \dfrac{dq}{dt} = \dfrac{0.5\rho v^2 S l C_m + P(y_P + x_P \sin\delta_P)}{I_{zz}}, \\ \dfrac{dh}{dt} = v\sin\theta \end{cases} \qquad (1)$$

where $(v,\theta,\alpha,q,h)$ are state variables, which initial values can vary within certain given limits. The expressions for aerodynamic coefficients are

$$\begin{cases} C_Y = C_{Y_\alpha}\sin(2\alpha) + C_{Y_{\delta_m}}\delta_m, \quad C_X = C_{X_0} + kC_Y^2, \\ C_m = C_{m_\alpha}\sin(2\alpha) + C_{m_{\delta_m}}\delta_m + C_{m_q}\dfrac{l}{v}q \end{cases}. \qquad (2)$$

Here $v$ is airspeed, $[m/s]$; $\rho$ is atmospheric density, $[\text{kg/m}^3]$; $S$ is wing area, $[\text{m}^2]$; $g$ is gravitational constant, $[\text{m/s}^2]$; $m$ is mass of the aircraft, $[\text{kg}]$; $h$ is flight altitude, $[\text{m}]$; $\alpha$ is angle of attack, $[\text{rad}]$; $\theta$ is flight path angle, $[\text{rad}]$; $\{C_Y, C_X, C_m\}$ are lift, drag and pitching moment aerodynamic coefficients of the aircraft respectfully; $I_{zz}$ is pitch axis aircraft mass moment of inertia, $[\text{kg}\times\text{m}^2]$; $q$ is pitch attitude angle rate of the aircraft, $[\text{rad/s}]$; $\vartheta$ is pitch attitude angle, $[\text{rad}]$; $t$ is time, $[\text{s}]$; $l$ is mean aerodynamic chord, $[\text{m}]$. The standard model of the atmosphere is $\rho = 1.2256(1 - 0.2257\cdot 10^{-4}\cdot h)^{4.256}$. The controls are

- $\delta_m \in \left[-\bar{\delta}_m; \bar{\delta}_m\right]$ in $[\text{rad}]$, aerodynamic control as the angular deflection of a movable horizontal stabilizer is at least continuous function of time;
- $\delta_P \in \left[-\bar{\delta}_P; \bar{\delta}_P\right]$ in rad, jet engine control as the angular deflection of a turbo-jet engine nozzle is at least continuous function of time, $[\text{rad}]$;
- $P \in \left[P_{\min}; P_{\max}\right]$ in N, aircraft turbojet engine thrust is a controlling parameter as a step-function taking on an admissible constant value within a given time segment. This kind of functions can also simulate engine failures.

The control aim is to find feedback control laws for $\delta_m$ and $\delta_P$ that make the aircraft automatically track the given time program of the flight path angle in the form of $\theta = \theta(t) = \theta_m \cdot [1 + \sin(\omega t)]$ with constant $\theta_m$ and, at that, the pitch attitude angle must be equal to some given constant value during the entire maneuver, namely $\vartheta = \vartheta' = const$, regardless of the values of atmospheric density, atmospheric perturbations, aerodynamic coefficients, the value of thrust, aircraft weight, etc.

The design of the tracking control will be done with the assumption that $C_{Y_{\delta_m}} \equiv 0$. This means that we will have used the simplified mathematical model to synthesize the control laws but we will utilize it for the original one (1) with $C_{Y_{\delta_m}} \neq 0$. This move definitely makes the problem of the control design much easier to solve. In fact, we cancel only one term, namely $C_{Y_{\delta_m}} \delta_m$, in the equations of the original model and the degree of complexity of the problem of designing the control drops dramatically. The computer simulation shows that the tactic does the trick via the wide-sense robustness provided by the designed control law that ensures mutual adequacy of the original and simplified mathematical models. By it we mean that the same control applied to both the approximate models of the thrust-vectored aircraft maneuvering produces the topologically equivalent phase spaces with quantitative differences being within certain imposed limits.

We will employ the same control design procedure, consisting of four stages, similar to the one used for the 2D system in Section 2. Four control law coefficients $\left(a_i < 0\right)_{i=1}^{i=4} \in \mathbb{R}^4$ will be introduced one-by-one through the stages.

Thus, first, we will use the simplified version of the original model governed by the system of equations (1) with

$$\left\{ C_Y = C_{Y_\alpha} \sin(2\alpha), \ C_X = C_{X_0} + kC_Y^2, \ C_m = C_{m_\alpha} \sin(2\alpha) + C_{m_{\delta_m}} \delta_m + C_{m_q} \frac{l}{v} q \right\} \quad (3)$$

instead of (2) simply by setting the coefficient $C_{Y_{\delta_m}}$ equal to zero in it. It will have its effect on the system (1) via the right-hand side $A_2(\bullet)$ of differential equation for $\theta$.

Second, we define our terminal manifolds, onto which the system (1)-(3) should be brought. They are the one-codimensional manifolds as follows

1. The first one is $M_\theta = \{\theta = \theta_m \cdot [1 + \sin(\omega t)]\}$.

2. The second one is $M_\vartheta = \{\vartheta = \vartheta'\}$, where $\vartheta' = const$.

The CLF $V = \bar{\theta}^2 + \bar{\vartheta}^2$ will ensure the asymptotical stability of CFDS (1)-(3) with respect to part of the variables $(\bar{\theta}, \bar{\vartheta}) = \bar{\varphi}(\bar{\varphi}^{-1})(\theta, \vartheta)$.

**Stage 1.** Forming the first terminal attractor that will represent the given time program of the flight path angle $\theta = \theta(t) = \theta_m \cdot [1 + \sin(\omega t)]$.

On this stage we turn the first one-codimensional terminal manifold $M_\theta$ into the attractor via the angle of attack $\alpha$ that will act as a controlling function on the flight path angle $\theta$ under the influence of the corresponding control laws for $\delta_p$ and $\delta_m$.

**Stage 1 - Step 1.** Changing the variable $\theta$ for the new one $\bar{\theta}$ in the system (1):

$$\theta = \bar{\theta} + \theta_m \cdot [1 + \sin(\omega t)], \quad \frac{d\theta}{dt} = \frac{d\bar{\theta}}{dt} + \theta_m \cdot \omega \cdot \cos(\omega t).$$

Thus, we obtain the differential equation for the new variable $\bar{\theta}$:

$$\frac{d\bar{\theta}}{dt} = \frac{P}{mv} \cdot \sin(\alpha + \delta_p) + \frac{0.5 \rho v S C_{Y_\alpha} \sin(2\alpha)}{m} -$$
$$-\frac{g}{v} \cos(\bar{\theta} + \theta_m \cdot [1 + \sin(\omega t)]) - \theta_m \cdot \omega \cdot \cos(\omega t).$$

**Stage 1 - Step 2.** Forming the right-hand side of the differential equation:

$$\frac{P}{mv} \cdot \sin(\alpha + \delta_p) + \frac{0.5 \rho v S C_{Y_\alpha} \sin(2\alpha)}{m} - \qquad (4)$$
$$-\frac{g}{v} \cos(\bar{\theta} + \theta_m \cdot [1 + \sin(\omega t)]) - \theta_m \cdot \omega \cdot \cos(\omega t) = a_1 \cdot \bar{\theta}.$$

**Stage 1 - Step 3.** Making the inverse change of the variable $\bar{\theta}$ back to $\theta$:

$$\bar{\theta} = \theta - \theta_m \cdot [1 + \sin(\omega t)].$$

We receive the following functional equation

$$A_2'(\bullet) - \theta_m \cdot \omega \cdot \cos(\omega t) - a_1 \cdot (\theta - \theta_m \cdot [1 + \sin(\omega t)]) = 0, \qquad (5)$$

where $A_2'(\bullet) = \frac{P}{mv} \sin(\alpha + \delta_p) + \frac{0.5 \rho v S C_{Y_\alpha} \sin(2\alpha)}{m} - \frac{g}{v} \cdot \cos\theta$.

This very important equation can be solved for the angle of attack $\alpha$ only by means of the numerical method. Let us designate its solution $\alpha = \varphi(t, P, m, v, \delta_p, h, \theta, \omega, a_1)$.

**Stage 2.** Forming the first intermediate attractor $M_\alpha$ defined by the expression (5) that will transfer the sought control to the terminal manifold $M_\theta$ through the angle of attack $\alpha$. At this stage we define the one-codimensional manifold

$$M_\alpha = \left\{ \begin{array}{l} A_2'(\bullet) - \theta_m \cdot \omega \cdot \cos(\omega t) - a_1 \cdot (\theta - \theta_m \cdot [1 + \sin(\omega t)]) = 0 \Leftrightarrow \\ \Leftrightarrow \alpha = \varphi(t, P, m, v, \delta_P, h, \theta, \omega, a_1) \end{array} \right\},$$

which is turned in the attractor by the corresponding control laws for $\delta_P$ and $\delta_m$.

**Stage 2 - Step 1.** Changing the phase variable $\alpha$ for the new one $\bar{\alpha}$:

$$\alpha = \bar{\alpha} + \varphi(t, P, m, v, \delta_P, h, \theta, \omega, a_1).$$

For the derivative we have: $\dfrac{d\alpha}{dt} = \dfrac{d\bar{\alpha}}{dt} + \dfrac{d\varphi(t, P, m, v, \delta_P, h, \theta, \omega, a_1)}{dt}$.

Thus, we obtain the differential equation for the new variable $\bar{\alpha}$:

$$\frac{d\bar{\alpha}}{dt} = q - \frac{P}{mv}\sin\left([\bar{\alpha} + \varphi(t, P, m, v, \delta_P, h, \theta, \omega, a_1)] + \delta_p\right) + \frac{g}{v}\cdot\cos\theta -$$

$$-\frac{d\varphi(t, P, m, v, \delta_P, h, \theta, \omega, a_1)}{dt} - \frac{0.5\rho v S C_{Y_\alpha} \sin(2[\bar{\alpha} + \varphi(t, P, m, v, \delta_P, h, \theta, \omega, a_1)])}{m}.$$

**Stage 2 - Step 2.** Forming the right-hand side of the differential equation:

$$q - \frac{P\sin([\bar{\alpha} + \varphi(\bullet)] + \delta_p)}{mv} - \frac{d\varphi(\bullet)}{dt} - \frac{\rho v S C_{Y_\alpha} \sin(2[\bar{\alpha} + \varphi(\bullet)])}{2m} + \frac{g\cos\theta}{v} = a_2 \bar{\alpha}. \quad (6)$$

**Stage 2 - Step 3.** Making the inverse change of the variable $\bar{\alpha}$:

$$\bar{\alpha} = \alpha - \varphi(t, P, m, v, \delta_P, h, \theta, \omega, a_1).$$

We receive the following functional equation:

$$q - A_2'(\bullet) - \frac{d\varphi(t, P, m, v, \delta_P, h, \theta, \omega, a_1)}{dt} = a_2\left[\alpha - \varphi(t, P, m, v, \delta_P, h, \theta, \omega, a_1)\right].$$

We can solve this equation for $q$. We have:

$$q = a_2\left[\alpha - \varphi(t, P, m, v, \delta_P, h, \theta, \omega, a_1)\right] + A_2'(\bullet) + \frac{d\varphi(t, P, m, v, \delta_P, h, \theta, \omega, a_1)}{dt}. \quad (7)$$

**Stage 3.** Forming the second terminal attractor $M_\vartheta$ that will represent the given time program of the pitch attitude angle $\vartheta$, namely $\vartheta = \vartheta'$.

At this stage we turn the second one-codimensional terminal manifold $M_\vartheta$ into the attractor with the help of the angular deflection of a turbojet engine nozzle $\delta_P$.

**Stage 3 - Step 1.** Changing the phase variable $\vartheta$ for the new one $\bar{\vartheta}$:

We have $\vartheta = \bar{\vartheta} + \vartheta'$ and its derivative is $\dfrac{d\vartheta}{dt} = \dfrac{d\bar{\vartheta}}{dt} = \dfrac{d\alpha}{dt} + \dfrac{d\theta}{dt}$.

We receive the following differential equation $\dfrac{d\bar{\vartheta}}{dt} = q$ for $\bar{\vartheta}$, where $q$ satisfies (7).

**Stage 3 - Step 2.** Forming the right-hand side of the differential equation:

$$q = a_4 \bar{\vartheta}. \quad (8)$$

**Stage 3 - Step 3.** Making the inverse change of the variable $\bar{\vartheta}$ back to $\vartheta$.

Let us return to the original phase variable $\vartheta$ through the expression $\bar{\vartheta} = \vartheta - \vartheta'$, which we plug in (8). Using (7) we obtain the functional equation:

$$a_4(\vartheta - \vartheta') = a_2[\alpha - \varphi(\cdot)] + A_2'(\cdot) + \frac{d\varphi(\cdot)}{dt}, \quad (9)$$

where $\dfrac{d\varphi(\cdot)}{dt} = \dfrac{\partial\varphi(\cdot)}{\partial t} + \dfrac{\partial\varphi(\cdot)}{\partial v}\dfrac{dv}{dt} + \dfrac{\partial\varphi(\cdot)}{\partial h}\dfrac{dh}{dt} + \dfrac{\partial\varphi(\cdot)}{\partial \theta}\dfrac{d\theta}{dt} + \dfrac{\partial\varphi(\cdot)}{\partial \delta_P}\dfrac{d\delta_P}{dt}$ and $\varphi(\cdot) = \varphi(t, P, m, v, \delta_P, h, \theta, \omega, a_1)$.

The equation (9) can be solved for $\dfrac{d\delta_P}{dt}$ and thus we obtain the control law for the angular deflection of a turbojet engine nozzle $\delta_P$ in the differential form:

$$\frac{d\delta_P}{dt} = \left\{\frac{\partial\varphi(\cdot)}{d\delta_P}\right\}^{-1} \times \left\{a_4(\vartheta - \vartheta') - a_2[\alpha - \varphi(\cdot)] - A_2'(\cdot) - W_0\right\}, \quad (10)$$

where $W_0 = \dfrac{\partial\varphi(\cdot)}{\partial t} + \dfrac{\partial\varphi(\cdot)}{\partial v}\dfrac{dv}{dt} + \dfrac{\partial\varphi(\cdot)}{\partial h}\dfrac{dh}{dt} + \dfrac{\partial\varphi(\cdot)}{\partial \theta}\dfrac{d\theta}{dt} =$

$$= \frac{\partial\varphi(\cdot)}{\partial t} + \frac{\partial\varphi(\cdot)}{\partial v}A_1(\cdot) + \frac{\partial\varphi(\cdot)}{\partial h}v\sin\theta + \frac{\partial\varphi(\cdot)}{\partial \theta}A_2'(\cdot).$$

**Stage 4.** Forming the second intermediate attractor $M_q$ that will transfer the sought control to the first terminal manifold $M_\theta$ through the phase variable $q$. The one-codimensional manifold $M_q$ is defined by the equation (7), namely:

$$M_q = \left\{q = a_2[\alpha - \varphi(\cdot)] + A_2'(\cdot) + \frac{d\varphi(\cdot)}{dt}\right\}.$$

**Stage 4 - Step 1.** Changing the phase variable $q$ for the new one $\bar{q}$:

$$q = \bar{q} + a_2[\alpha - \varphi(\cdot)] + A_2'(\cdot) + \frac{d\varphi(\cdot)}{dt}.$$

For the derivative we have:

$$\frac{dq}{dt} = \frac{d\bar{q}}{dt} + a_2\left(\frac{d\alpha}{dt} - \frac{d\varphi(\cdot)}{dt}\right) + \frac{dA_2'(\cdot)}{dt} + \frac{d^2\varphi(\cdot)}{dt^2}.$$

The new differential equation for $\bar{q}$ has the following look:

$$\frac{d\bar{q}}{dt} = \frac{1}{I_{zz}}\left\{0.5\rho v^2 Sl\left[C_{m_\alpha}\sin(2\alpha) + C_{m_{\delta_m}}\delta_m + W_1\right] + P(y_P + x_P\sin\delta_P)\right\} -$$

$$- a_2\left(\frac{d\alpha}{dt} - \frac{d\varphi(\cdot)}{dt}\right) - \frac{dA_2'(\cdot)}{dt} - \frac{d^2\varphi(\cdot)}{dt^2},$$

where $W_1 = C_{m_q}\dfrac{l}{v}\left(\bar{q} + a_2(\alpha - \varphi(\cdot)) + A_2'(\cdot) + \dfrac{d\varphi(\cdot)}{dt}\right)$.

**Stage 4 - Step 2.** Forming the right-hand side of the differential equation:

$$\frac{1}{I_{zz}}\left\{0.5\rho v^2 Sl\left[C_{m_\alpha}\sin(2\alpha)+C_{m_{\delta_m}}\delta_m+W_1\right]+P(y_P+x_P\sin\delta_P)\right\}- $$
$$-a_2\left(\frac{d\alpha}{dt}-\frac{d\varphi(\cdot)}{dt}\right)-\frac{dA_2'(\cdot)}{dt}-\frac{d^2\varphi(\cdot)}{dt^2}=a_3\bar{q}. \quad (11)$$

**Stage 4 - Step 3.** Making the inverse change of the variable $\bar{q}$ back to $q$.

To return to the initial phase variable $q$ we use the following expression:

$$\bar{q}=q-a_2[\alpha-\varphi(\cdot)]-A_2'(\cdot)-\frac{d\varphi(\cdot)}{dt}.$$

We have the functional equation

$$\frac{1}{I_{zz}}\left\{0.5\rho v^2 Sl\left[C_{m_\alpha}\sin(2\alpha)+C_{m_{\delta_m}}\delta_m+C_{m_q}\frac{l}{v}q\right]+P(y_P+x_P\sin\delta_P)\right\}-$$
$$-a_2\left(\frac{d\alpha}{dt}-\frac{d\varphi(\cdot)}{dt}\right)-\frac{dA_2'(\cdot)}{dt}-\frac{d^2\varphi(\cdot)}{dt^2}=a_3\left\{q-a_2[\alpha-\varphi(\cdot)]-A_2'(\cdot)-\frac{d\varphi(\cdot)}{dt}\right\}.$$

Solving this equation for $\delta_m$ and we get the control law for the aerodynamic control:

$$\delta_m=\frac{1}{C_{m_{\delta_m}}}\left\{\frac{1}{0.5\rho v^2 Sl}\left[I_{zz}(W_2+W_3)-P(y_P+x_P\sin\delta_P)\right]-W_4\right\}, \quad (12)$$

where $W_2=a_3\left\{q-a_2[\alpha-\varphi(\cdot)]-A_2'(\cdot)-\frac{d\varphi(\cdot)}{dt}\right\}$, $W_4=C_{m_\alpha}\sin(2\alpha)+C_{m_q}\frac{l}{v}q$,

$$W_3=a_2\left(\frac{d\alpha}{dt}-\frac{d\varphi(\cdot)}{dt}\right)+\frac{dA_2'(\cdot)}{dt}+\frac{d^2\varphi(\cdot)}{dt^2}.$$

Ultimately, the expressions (10) and (12) present the wide-sense robust tracking feedback control laws for $\delta_P$ and $\delta_m$ respectively making $M_\theta$ and $M_\vartheta$ the stable invariant manifolds for system (1)-(2) and the asymptotically stable invariant manifolds or attractors for system (1)-(3) with CLF $V=\bar{\theta}^2+\bar{\vartheta}^2$. The control aim is attained through two intermediate attractors $M_\alpha$ and $M_q$ also created by $\delta_P$ and $\delta_m$. The results of the computer simulation and its analysis can be found in [10].

## 4 Conclusion

The canonical form of Lyapunov's second method requires the manipulations with the right-hand sides of ordinary differential equations as the starting point of the procedure. This fact arms the researcher with the opportunity to profit if they include controlling functions. The latter are the certain degrees of freedom to be handled of the researcher's choice. Knowledge of first integrals is a rare bird for nonlinear equations. And this is a huge stumbling block to proceed. But the controlling functions, which are free to be chosen within a given constrains, practically annihilate this impedance.

This fact allows us to design two sought wide-sense robust and stable in the large tracking control laws. Their structure organized as the two-stage cascade of four attractors is given below.

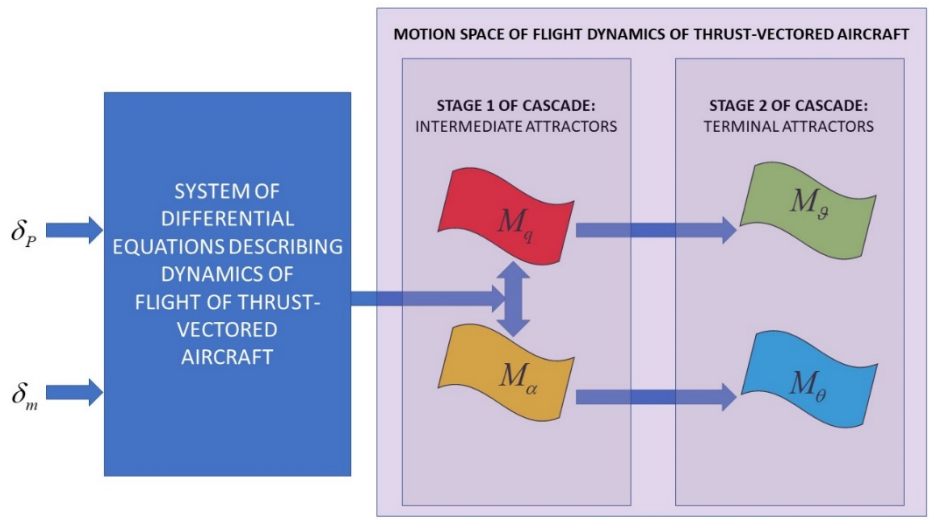

**Fig. 3.** The two-stage cascade of four attractors embedded in the motion space of the 5D nonlinear mathematical model of the maneuvering of the thrust-vectored aircraft being under the designed tracking control laws for the angular deflection of a turbojet engine nozzle (10) and the angular deflection of a movable horizontal stabilizer (12) that ensure wide-sense robustness and stability in the large of the flight dynamics.

## References

1. Martynyuk-Chernienko, Y.A. Application of the Canonical Lyapunov Function in the Theory of Stability of Uncertain Systems. International Applied Mechanics 36, 1112–1118 (2000). https://doi.org/10.1023/A:1026621319709.
2. Schwartz, Carla A., Yan, A. Construction of Lyapunov Functions for Nonlinear Systems Using Normal Forms. Journal of Mathematical Analysis and Applications. 216, 521-535, 1997.
3. Ruo-Shi Yuan, Yi-An Ma, Bo Yuan, Ping Ao. Lyapunov function as potential function: A dynamical equivalence. 2013 Chinese Phys. B 23 010505.
4. Galperin, E.A. Some Generalizations of Lyapunov's Approach to Stability and Control. Nonlinear Dynamics and Systems Theory, 2(1) (2002) 1–23.
5. Szydłowski, M., Krawiec, A. Lyapunov function for cosmological dynamical system. Demonstratio Mathematica, DOI: https://doi.org/10.1515/dema-2017-0005.
6. Giesl, P., Hafstein, S. Review on computational methods for Lyapunov functions. Discrete & Continuous Dynamical Systems - B, 2015, 20 (8) : 2291-2331. doi: 10.3934/dcdsb.2015.20.2291.
7. Dos Santos, V., Bastin, G., Coron, J-M., d'Andréa-Novel, B. Boundary control of systems of conservation laws: Lyapunov stability with integral actions. IFAC Proceedings Vol-

umes, Volume 40, Issue 12, 2007, Pages 312-317. https://doi.org/10.3182/20070822-3-ZA-2920.00052

8. Sparavalo, M. K. The Lyapunov Concept of Stability from the Standpoint of Poincare Approach: General Procedure of Utilization of Lyapunov Functions for Non-Linear Non-Autonomous Parametric Differential Inclusions, 2014. arXiv:1403.5761v4 [cs.SY].
9. Sparavalo, M. K. A method of goal-oriented formation of the local topological structure of co-dimension one foliations for dynamic systems with control. J. Automat. Inform. Sci. (1992), 25(5), 65-71 (1993).
10. Sparavalo, M. K. Adequate Mathematical Modelling by Wide-sense Robust Control Design in a Thrust-Vectored Flight Dynamics Problem. CEAS Aeronaut J **11,** 289–301 (2020). https://doi.org/10.1007/s13272-019-00425-x.